\documentclass{aa}
\usepackage{graphics}

\begin{document}      
   
   \title{The characteristic polarized radio continuum distribution of cluster spiral galaxies}

   \author{B.~Vollmer\inst{1}, M.~Soida\inst{2}, R.~Beck\inst{3},
     M.~Urbanik\inst{2}, K.T.~Chy\.zy\inst{2}, K.~Otmianowska-Mazur\inst{2}, 
     J.D.P.~Kenney\inst{4}, J.H.~van~Gorkom\inst{5}}

   \offprints{B.~Vollmer: bvollmer@astro.u-strasbg.fr}

   \institute{CDS, Observatoire astronomique de Strasbourg, 11, rue de l'universit\'e,
	      67000 Strasbourg, France \and
	      Astronomical Observatory, Jagiellonian University,
	      Krak\'ow, Poland \and
	      Max-Planck-Insitut f\"{u}r Radioastronomie, Auf dem H\"{u}gel 69, 53121 Bonn, Germany \and
	      Yale University Astronomy Department, P.O. Box 208101, New Haven, CT 06520-8101, USA \and
	      Department of Astronomy, Columbia University, 538 West 120th Street, New York, 
	      NY 10027, USA
              }

   \date{Received / Accepted}

   \authorrunning{Vollmer et al.}
   \titlerunning{VLA polarized radio continuum observations of Virgo spirals}

\abstract{
Deep observations of 6~cm polarized radio continuum emission of 8 Virgo spiral galaxies are presented.
All galaxies show strongly asymmetric distributions of polarized intensity with elongated ridges
located in the outer galactic disk. Such features are not found in existing observations
of polarized radio continuum emission of field spiral galaxies, where the distribution
of 6~cm polarized intensity is generally relatively symmetric and strongest in the interarm regions. 
We therefore conclude that most Virgo spiral galaxies
and most probably the majority of cluster spiral galaxies show asymmetric distributions of polarized
radio continuum emission due to their interaction with the cluster environment.  
The polarized continuum emission is sensitive to compression and shear motions in the plane of the sky and thus 
contains important information about velocity distortions caused by these interactions.
\keywords{Galaxies: interactions -- Galaxies: ISM -- Galaxies: kinematics and dynamics -- Galaxies: magnetic fields -- Galaxies: clusters@ individual: Virgo -- Radio continuum: galaxies}}

\maketitle

\section{Introduction \label{sec:introduction}}

Cluster spiral galaxies undergo different interactions
which can change their gas distribution and content, stellar distribution, and
morphology. These interactions can be tidal (with the cluster potential:
Byrd \& Valtonen 1990, Valluri 1999 or galaxy--galaxy interactions: galaxy ``harassment''; 
Moore et al. 1998) or between the galaxy's ISM and the hot intracluster gas 
(ram pressure stripping: Gunn \& Gott 1972). To determine the nature of an interaction,
multi wavelength observations are required. A tidal interaction leads to a distorted
stellar and gas distribution observable in the optical (stars; gas: H$\alpha$) and the radio (gas: CO, HI).
Spectroscopic observations of the ISM have the advantage to provide the distribution of
the velocities along the line of sight. 
However, no direct information about the velocity
components in the plane of the sky are accessible in this way.
The distribution of polarized radio continuum emission can provide this information
(Beck 2005).

Polarized radio continuum emission is due to relativistic electrons with density $n_{\rm e}$
gyrating around the regularly oriented large scale magnetic field $B$:
$S_{\rm PI} \propto n_{\rm e} B^{2}$. The polarized radio continuum emission is enhanced in 
regions where shear and compression of regular or random magnetic fields occur. 
Polarized emission reveals compression regions much better than any other tracer,
even outside starforming regions.
From spectroscopic observations non-circular motions of the order of $\sim$10~km\,s$^{-1}$ 
induced by an interaction can 
be determined by a detailed analysis of a galaxy's velocity field (e.g. Schoenemakers et al. 1997).
On the other hand, the distribution of polarized radio continuum emission represents a very
sensitive tool to uncover transverse motions of the ISM (Beck 2005) even in the case of
unfavorable inclinations (close to face-on).
Therefore, the information contained in polarized radio continuum emission is 
complementary to that of H$\alpha$, CO, and H{\sc i} observations.
The total radio continuum emission is sensitive to the turbulent small scale magnetic 
field which is usually a factor of 2--5 larger than the regular large scale magnetic field
in spiral arms and 1--2 times larger in the interarm regions at a typical resolution of a few 100~pc
(Beck 2001).
Whenever there is enhanced turbulence due to an enhanced star formation efficiency, the
large scale magnetic field is diminished. The polarized radio continuum emission has to be
observed at a frequency high enough to avoid significant Faraday rotation (typically $>2$~GHz).

So far, only 2 Virgo spiral galaxies have been imaged with the VLA in polarized radio
continuum emission, because this is very time consuming. 
In NGC~4254 an asymmetric pattern of magnetic spiral arms was found at 3 and 6~cm
with the strongest polarization in the south, outside the optical arm 
(Chy\.zy et al. 2006).
In NGC~4522 Vollmer et al. (2004) also discovered an asymmetric distribution of the
6~cm polarized emission located at the eastern edge of the galactic disk, opposite to the 
western extraplanar H{\sc i} gas which is pushed out of the galactic plane by ram pressure.
Using the Effelsberg 100m telescope Soida et al. (1996) and We\.zgowiec et al. (2007)
found asymmetric distributions of polarized radio continuum emission at 3 and 6~cm 
in NGC~4254, NGC~4438, NGC~4501, NGC~4535, and NGC~4654.
Encouraged by these results we undertook a first high resolution VLA survey of 8 Virgo spiral galaxies
in polarized radio continuum emission at 6 and 20~cm. These galaxies span an order of magnitude 
in luminosity from 0.2 to 2~L$^{*}$, have different inclinations, and 
are located at various distances between 1$^{\circ}$ and 4$^{\circ}$ from M~87.
In this letter we report on the
spectacular results of the observations of 6~cm polarized radio continuum emission and we show that its
distribution is different in cluster and field spiral galaxies.

\section{Observations\label{sec:observations}}

The 8 Virgo spiral galaxies where observed between November 8th 2005 and January 10th 2006
with the Very Large Array (VLA) of the National Radio Astronomy Observatory
(NRAO)\footnote{NRAO is a facility of National Science Foundation
operated under cooperative agreement by Associated Universities, Inc.}
in the D array configuration. The band passes were $2\times 50$~MHz.
We used 3C286 as the flux calibrator and 1254+116 as the phase calibrator, the latter of
which was observed every 40~min. 
Maps where made for both wavelengths using the AIPS task IMAGR with ROBUST=3.
The final cleaned maps were convolved to a beam size of $18'' \times 18''$.
The bright radio source M87 caused sidelobe effects enhancing the rms noise level of NGC~4438.
We ended up with an rms level of the linear polarization, taken to be the mean rms
in Stokes Q and U, between $9$ and $14$~$\mu$Jy/beam (Table~\ref{tab:table}).
\begin{table}
      \caption{Integration times and rms.}
         \label{tab:table}
      \[
         \begin{array}{llcccc}
           \hline
           \noalign{\smallskip}
           {\rm galaxy\ name } & {\rm m_{\rm B}^{(1)}} & i^{(2)} & {\rm Dist.^{(3)}} &{\rm integration} & {\rm rms}  \\
	    &  & & & {\rm time} & (\mu{\rm Jy/} \\
	    & {\rm (mag)} & {\rm (deg)}& {\rm (deg)} & {\rm (h:min)} & {\rm beam)} \\
	   \noalign{\smallskip}
	   \hline
	   \noalign{\smallskip}
	   {\bf NGC~4321} & 10.02 & 27^{\rm (a)} & 4.0 & 7:45 & 9 \\ 
           \noalign{\smallskip}
	   \hline
	   \noalign{\smallskip}
	   {\bf NGC~4388} & 11.87 & 77^{\rm (a)} & 1.3 & 9:25 & 9  \\
	   \noalign{\smallskip}
           \hline
	   \noalign{\smallskip}
	   {\bf NGC~4396} & 13.07 & 72^{\rm (b)} & 3.5 & 8:00 & 9  \\
	   \noalign{\smallskip}
           \hline
	   \noalign{\smallskip}
	   {\bf NGC~4402} & 12.64 & 74^{\rm (c)} & 1.4 & 5:00 & 13  \\
	   \noalign{\smallskip}
           \hline
	   \noalign{\smallskip}
	   {\bf NGC~4438} & 11.12 & 68^{\rm (a)}/85^{\rm (c)} & 1.0 & 7:45 & 14  \\
	   \noalign{\smallskip}
           \hline
	   \noalign{\smallskip}
	   {\bf NGC~4501} & 10.50 & 57^{\rm (a)} & 2.0 & 3:55 & 11  \\
	   \noalign{\smallskip}
           \hline
	   \noalign{\smallskip}
	   {\bf NGC~4535} & 10.73 & 43^{\rm (a)} & 4.3 & 9:00 & 9  \\
	   \noalign{\smallskip}
           \hline
	   \noalign{\smallskip}
	   {\bf NGC~4654} & 11.31 & 51^{\rm (a)} & 3.4 & 7:50 & 9  \\
	   \noalign{\smallskip}
           \hline
        \end{array}
      \]
\begin{list}{}{}
\item[$^{(1)}$ This research has made use of the GOLD Mine]
\item[\ \ \ \ Database (Gavazzi et al. 2003)]
\item[$^{(2)}$ Inclination angle; $^{(3)}$ Distance from M~87] 
\item[$^{(a)}$ from Cayatte et al. (1990); $^{(b)}$ from NED]
\item[$^{(c)}$ from Kenney et al. (1995)]
\end{list}
\end{table}

\section{Results}

\begin{figure*}
\centering
\resizebox{13cm}{!}{\includegraphics{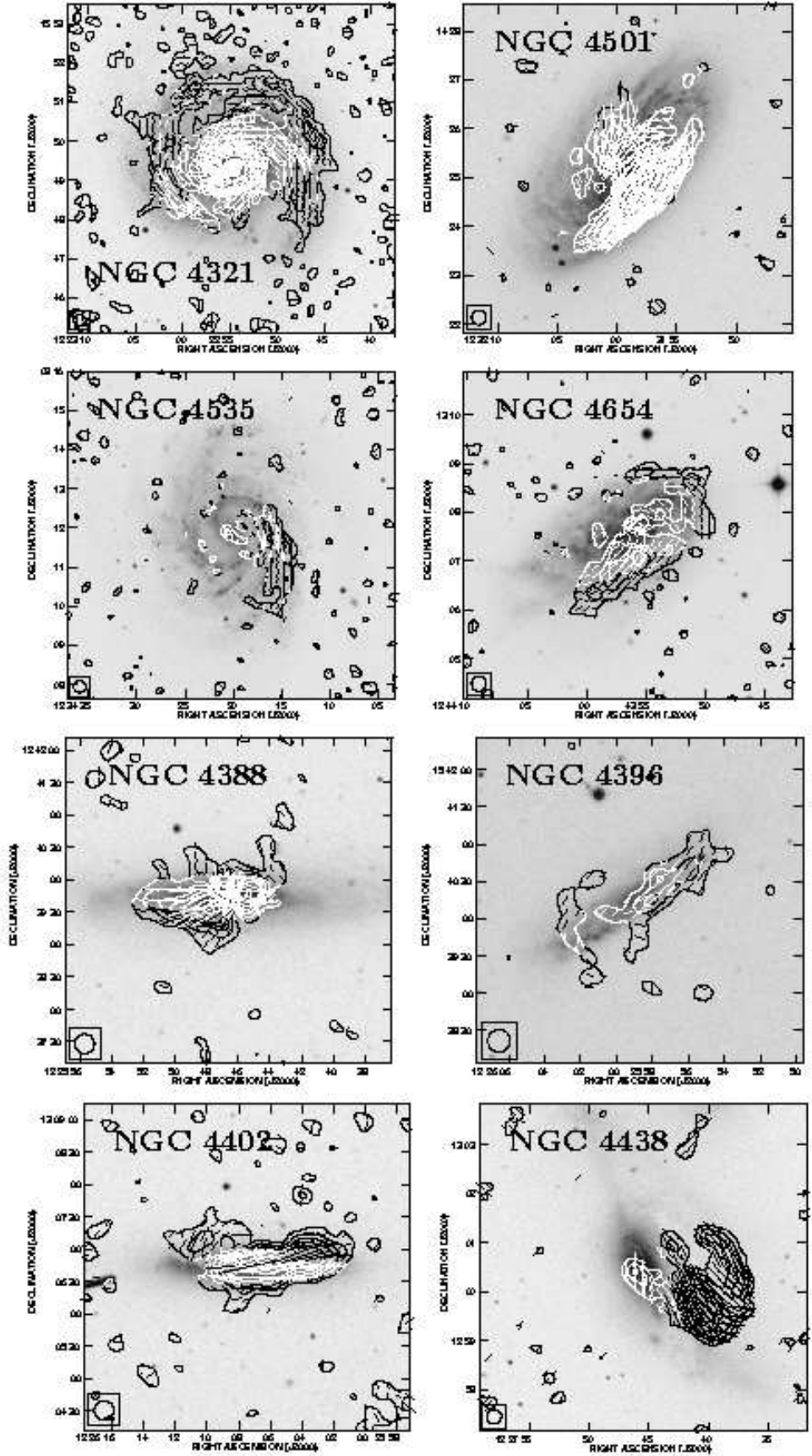}}
\caption{6~cm polarized intensities as contours and the 
	vectors of the magnetic field uncorrected for
	Faraday rotation on DSS B band images. 
	The size of the lines is proportional to the
	intensity of the polarized emission. The contour levels are
	(4, 8, 12, 16, 20, 30, 40, 50, 100, 150)$\times \eta$~$\mu$Jy
	($\eta = 8$ for NGC~4388, NGC~4396, NGC~4402, NGC~4654; 
	$\eta = 10$ for NGC~4321 and  NGC~4535; $\eta = 15$ for
	NGC~4438, NGC~4501).
	The beam ($18'' \times 18''$) is plotted in the lower left
	corner of the image.
}
\label{fig:pi}
\end{figure*}

The resulting maps of the 6~cm polarized radio continuum emission are shown as 
contours overlaid on optical B-band DSS images in Fig.~\ref{fig:pi}. The short lines
delineate the orientation of the magnetic field uncorrected for Faraday rotation
(hereafter refereed to as ``B-vectors'').
Their length is proportional to the polarized intensity.
In the 6~cm Effelsberg observations of NGC~4438, NGC~4501,
NGC~4535, and NGC~4654 We\.zgowiec et al. (2007) found a maximum Faraday rotation of 60~rad/m$^{2}$. 
Therefore, the B-vectors represent the sky-projected regular field to an accuracy
of $\pm 13^{\circ}$.
We divide our sample into highly and mildly inclined spiral galaxies.
Only the most distant galaxy from the cluster center, {\it NGC~4321}, 
shows the characteristic distribution of an unperturbed field spiral galaxy:
the polarized radio continuum emission is mainly found in regions between the spiral arms, 
because turbulence linked to star formation in the spiral arms destroys the large scale magnetic field
(see Sec.~\ref{sec:introduction}). There is, however, a north-south asymmetry which is 
also visible in the optical and H{\sc i} images. 
All other Virgo spiral galaxies show a pronounced asymmetric
distribution of the 6~cm polarized radio continuum emission with ridges at the outer
part of the galactic disks:\\
In {\it NGC~4501} we observe a strongly asymmetric distribution of polarized intensity
with an extended maximum in the outer south western part of the disk which coincides
with a region of high column density H{\sc i} (Cayatte et al. 1990). There is also
a small polarized spot in the opposite part of the disk.
{\it NGC~4535} has an overall low emission of polarized radio continuum emission.
The only region where we detect polarized emission is in the center and the south west of the 
outer galactic disk located on the ring-like H{\sc i} distribution.\\
{\it NGC~4654} shows an extended ridge of polarized emission in the southern part of the 
outer disk. We also find emission at the western and northern outer rim of the galactic disk.
We do not detect any polarized emission in the south eastern extended H{\sc i} tail (Phookun \& Mundy 1995).\\
{\it NGC~4396} is a faint edge-on galaxy showing a ridge of polarized radio continuum
emission in the north western part of the outer disk. Chung et al. (2007 in prep.) have
detected an H{\sc i} tail on this side of the galaxy extending to the north west.\\
In {\it NGC~4402} we observe a strong maximum of polarized
intensity in the western part of the disk where Crowl et al. (2005) detected 
extraplanar atomic hydrogen. There is no detection in the eastern outer disk.
In addition, we observe extraplanar polarized emission above the disk plane in the 
north east of the galactic disk, which coincides with the extended 20cm radio continuum
emission observed by Crowl et al. (2005). The observed H{\sc i} and 20~cm continuum
asymmetries are caused by ram pressure stripping.\\
{\it NGC~4388} shows a ridge of polarized radio continuum emission in the south eastern part
of the nearly edge-on disk. We observe a decreased polarized intensity to the south west
and north east of the galaxy center. This is due to beam depolarization 
caused by the nuclear outflow which is visible in H$\alpha$ emission (Veilleux et al. 1999). 
Thus, it is not excluded that the south eastern
ridge of polarized intensity continues towards the west and then turns to the north.\\
{\it NGC~4438} has strong tidal arms and shows the most surprising distribution of
polarized radio continuum emission. Within the disk/bulge region we observe a maximum to
the south east. However, most of the polarized emission is extraplanar and coincides 
with detections of extraplanar CO (Combes et al. 1988, Vollmer et al. 2005), H{\sc i}
(Cayatte et al. 1990) and H$\alpha$ (Kenney et al. 1995).

\section{Discussion}

All 8 observed Virgo spiral galaxies show a strongly asymmetric distribution of polarized 
radio continuum emission with elongated ridges located in the outer galactic disk.
Such polarized emission ridges are also observed in two other Virgo
spiral galaxies: NGC~4254 (Chy\.zy et al. 2006) and NGC~4522 (Vollmer et al. 2004).
To investigate if this kind of distribution is characteristic for
Virgo cluster spiral galaxies, we will review the distribution of polarized
radio continuum emission in field spiral galaxies. Since we have 4 edge-on galaxies
in our sample, we will first discuss results on highly inclined field spiral galaxies. 

Polarized radio continuum emission data at 20~cm of edge-on galaxies are available, e.g. 
for NGC~4631 (Hummel et al. 1988), NGC~891 (Hummel et al. 1991), NGC~4565 (Sukumar \& Allen 1991),
and NGC~5907 (Dumke et al. 2000). Whereas NGC~4631, NGC~891, and NGC~4565
show fairly symmetric distributions of polarized emission, NGC~5907 
has a polarized ridge in the south-west. However, one has to take into account
that these data suffer from severe depolarization, because
of the long wavelength and the long line of sight through the whole galactic disk. 
Thus one mainly sees the part of the galaxy which is closer to the observer.
This effect already causes an asymmetry along the minor axis. Spiral
arms can then cause asymmetries along the major axis. Therefore,
it is necessary to observe these galaxies at shorter wavelengths, where
depolarization is small. Sukumar \& Allen (1991) and T\"{u}llmann et al. (2000)
observed the nearby edge-on galaxies NGC~891 and NGC~5775 at 6~cm with the VLA. They found 
fairly symmetric distributions of polarized emission. 
Dumke et al. (2000) observed NGC~5907 at 6~cm
in polarization. They found that, despite the asymmetric distribution of the
polarized emission at 20~cm, the 6~cm polarized emission distribution is
almost symmetric. 
Thus, we conclude that highly inclined field galaxies can have heavily asymmetric
polarized emission distributions at 20~cm caused by depolarization, but not at 6~cm.

A compilation of imaging polarization measurements of 17 nearby galaxies 
at 6~cm with the VLA, ATCA, and the Effelsberg 100m telescope is available on
the website of the Max-Planck-Institut f\"{u}r Radioastronomie 
Bonn\footnote{http://www.mpifr-bonn.mpg.de/staff/wsherwood/mag-fields.html}.
At 6~cm wavelength, none of the non-interacting, isolated galaxies (NGC~2276 has a companion,
M~101 is a member of a group) shows prominent maxima or ridges 
in the polarized radio continuum emission at the outer edge of the galactic disk as
it is observed in the extended sample of 10 Virgo spiral galaxies.
We conclude that these maxima are characteristic for cluster and probably group
spiral galaxies. 

Since all asymmetric emission ridges are located in the outer galactic disks,
they are most probably due to external influences of the cluster environment
on the galaxies. These influences can be of tidal or hydrodynamic nature
(see Sect.~\ref{sec:introduction}). Thus the distribution of polarized
radio continuum emission contains important information about
velocity distortions caused by the interaction of a spiral galaxy with its cluster environment.
In addition, this information is complementary to that of CO or H{\sc i} data
cubes, because shear or compression motions can have dominant velocity components
in the plane of the sky. 
It is very difficult, if not impossible, to predict the distribution of polarized
radio continuum emission on the basis of CO or H{\sc i} observations, because of
the complex evolution of the magnetic field (induction equation) and beam
depolarization effects. Therefore, it is necessary to make detailed MHD modelling 
of individual galaxies for direct comparison with observations.
In Otmianowska-Mazur \& Vollmer (2003) we demonstrated
the feasibility of the method and in Vollmer et al. (2004) and Soida et al. (2006) 
we applied it successfully to the Virgo spiral galaxies NGC~4522 and NGC~4654, respectively.
In both cases the comparison between the observed and the modelled polarized
radio continuum emission distribution confirm the scenarios based on detailed
numerical simulations and deep H{\sc i} observations.

Based on a sample of 10 Virgo spiral galaxies observed with the VLA we conclude that the 6~cm polarized 
radio continuum emission distribution in Virgo spiral galaxies (this paper) is different from that
of field spiral galaxies (Beck 2005), i.e. the distribution is strongly asymmetric with elongated ridges
in the outer part of the galactic disks. These data contain important information
on the influence of the cluster environment on a spiral galaxy. Quantitative information
can be extracted by a detailed comparison with MHD models including tidal and
hydrodynamic (ram pressure) interactions.

\begin{acknowledgements}
This work was supported by Polish-French (ASTRO-LEA-PF) cooperation program,
and by Polish Ministry of Sciences and Higher Education grant PB 378/P03/28/2005
and 2693/H03/2006/31.
\end{acknowledgements}

\end{document}